\documentclass{article}

\usepackage{arxiv}

\usepackage[utf8]{inputenc} % allow utf-8 input
\usepackage[T1]{fontenc}    % use 8-bit T1 fonts
\usepackage{hyperref}       % hyperlinks
\usepackage{url}            % simple URL typesetting
\usepackage{booktabs}       % professional-quality tables
\usepackage{amsfonts}       % blackboard math symbols
\usepackage{nicefrac}       % compact symbols for 1/2, etc.
\usepackage{microtype}      % microtypography
\usepackage{lipsum}		% Can be removed after putting your text content
\usepackage{graphicx}
\usepackage{amsmath}
\usepackage{bm}
\usepackage{amssymb}
\usepackage{multicol}
\usepackage{float}

\title{Real-Time Time-Dependent Density Functional Theory within FHI-aims}

\author{Joscha~Hekele \\
	Faculty of Physics\\
	University of Duisburg-Essen\\
	Lotharstr. 1, 47057 Duisburg \\
	\texttt{joscha.hekele@uni-due.de} \\
	\And
	Peter Kratzer \\
	Faculty of Physics\\
	University of Duisburg-Essen\\
	Lotharstr. 1, 47057 Duisburg\\
	\texttt{peter.kratzer@uni-due.de} \\
}

\date{}

\renewcommand{\headeright}{}
\renewcommand{\undertitle}{}

\hypersetup{
pdftitle={},
pdfsubject={},
pdfauthor={Joscha~Hekele, Peter~Kratzer},
pdfkeywords={DFT, Real-Time TDDFT, First Principles, Theoretical Spectroscopy},
}

\begin{document}
\maketitle

\begin{abstract}
	Real-Time Time-Dependent Density Functional Theory (TDDFT) has become an attractive tool to model quantum dynamics on a first-principles Density Functional Theory level. In recent years, several developments and applications in this field were achieved and hopefully lead to new insights. We present here our versatile and efficient Real-Time TDDFT implementation into the all-electron numerical basis-set DFT code package FHI-aims. This article is meant as a short overview on how we performed this task and what can be done with our implementation. We further shed light on the connection of the basis set size to the accuracy of absorption spectrum simulation results.
\end{abstract}

% keywords can be removed

\begin{multicols*}{2}

% 3-7, 8, 9, 10, 11, 16-26, 27-29

\section{Introduction}
Real-Time Time-Dependent Density Functional Theory (RT-TDDFT) approaches \cite{tddft-fund} have gained more and more interest in the past two decades, where both the number of numerical studies and the introduction of computer codes incorporating this feature denote this trend \cite{elktddft, siestatddft, octopustddft}. Especially increasingly efficient Density Functional Theory (DFT) implementations enable the possibly costly real-time propagation of single-particle electronic states. \\
In contrast to the more widely used Linear-Response TDDFT (LR-TDDFT) approach \cite{casidalr}, RT-TDDFT is also able to describe the non-linear response of an electronic system, making it one of the few feasible tools to do so. Typical applications include simulations of laser-matter interaction with intense laser pulses \cite{tddft-tsolakidis, tddft-meng, tddft-aikens} or non-adiabatically coupled electron-ion motion in the framework of Ehrenfest Dynamics (ED) \cite{ehrenfest-base, ehrenfest-li, ehrenfest-tully, ehrenfest-isborn, ehrenfest-gpaw}.\\
While being conceptually rather simple, technical implications for an actual implementation and also practical usability can be more demanding since the numerical integration of time-dependent Kohn-Sham wavefunctions enforces several numerical considerations. It is thus also our aim to provide comprehensive insight into our implementation technique, formal implications and usability. We chose the numeric atom-centered basis function based all-electron real-space DFT code package FHI-aims \cite{fhiaims} to profit from its rich functionality and numerical efficiency to pave the way for a modern and versatile RT-TDDFT code\footnote{The code can be obtained from \url{https://aimsclub.fhi-berlin.mpg.de}}.\\
This paper is structured as follows: we first give a short overview of the theoretical fundament after which some brief implementation details will be presented. 
%After this, we provide a short tutorial-like overview on what can be done with our code and how. 
To show the validity of our implementation, we additionally present numerical and physical results for a standard test system -- in this frame, we try to give more insight about the influence of the chosen basis set and selected numerical parameters onto the simulation accuracy. In addition, we discuss the computational scalability with regard to basis set size.

\section{Theoretical Basis of RT-TDDFT}
\label{sec:theory}
Time-Dependent DFT is theoretically funded on the theorems of Runge and Gross \cite{rungegross} and Van Leeuwen \cite{vanleeuwen}, enabling the DFT framework to be used for the time-dependent electron density $\rho(\mathbf{r},t)$. The solution of the time-dependent Kohn-Sham (KS) equation,
\begin{equation}
i\frac{\partial}{\partial t}\psi^\mathrm{KS}_n(\mathbf{r},t) = \mathcal{H}^\mathrm{KS}[\rho(t),t]\psi_n^\mathrm{KS}(\mathbf{r},t),
\end{equation}
describes the time evolution of the electronic Kohn-Sham single-particle orbitals $\psi^\mathrm{KS}_n(\mathbf{r},t)$ (we use atomic units in this article). The KS Hamiltonian $\mathcal{H}^\mathrm{KS}[\rho(t),t]$ is time-dependent via the electron density but also possibly via a time-dependent external potential. It is given as
\begin{align}
\mathcal{H}^\mathrm{KS}[\rho(t),t] &= \mathcal{T}_\mathrm{el} + \mathcal{V}_\mathrm{ext}(t) + \mathcal{V}_H[\rho(t)] + \mathcal{V}_\mathrm{XC}[\rho(t)]
\end{align}
where $\mathcal{T}_\mathrm{el}$ is the electronic kinetic operator, $\mathcal{V}_\mathrm{ext}(t)$ is the external potential, $\mathcal{V}_H[\rho(t)]$ is the Hartree potential and $\mathcal{V}_\mathrm{XC}[\rho(t)]$ is the exchange-correlation potential. It can be seen that an explicit time-dependence is imposed by the external potential, e.g. a laser field or a dynamical ionic potential, and an implicit time-dependence by the functional dependence of the Hartree and XC potentials on the time-dependent electron density.\\
For our next discussion, we define the external potential as
\begin{align}
\mathcal{V}_\mathrm{ext}(t) &= \mathcal{V}_\mathrm{ion}(t) + \mathcal{V}_\mathrm{field}(t)
\end{align}
where $\mathcal{V}_\mathrm{ion}(t)$ denotes an ionic potential and $\mathcal{V}_\mathrm{field}(t)$ the potential of a possible external electric field. The interaction of electrons with an external electric field is usually described by the length gauge
\begin{align}
\mathcal{V}_\mathrm{field}(t) &= \mathbf{r}\cdot\mathbf{E}(t)
\end{align}
or the velocity gauge 
\begin{align}
\mathcal{V}_\mathrm{field}(t) &= i\mathbf{A}(t) + \frac{1}{2}\mathbf{A}^2(t)
\end{align}
where the electric field $\mathbf{E}(t)=-\partial_t \mathbf{A}(t)$ and the vector potential are here given in the dipole approximation, neglecting any spatial dependence of the electric field. 

% Ehrenfest dynamcis / other theoretical details?

\section{Implementation}
The FHI-aims code \cite{fhiaims} is based on numerical atom-centered basis functions in an all-electron description. The single-particle KS orbitals are thus expressed as a linear combination of non-orthogonal numerical atom-centered basis functions $\phi_i \in \mathbb{R}$, each associated with a corresponding atom $I$:
\begin{align}
\psi_n^\mathrm{KS}(\mathbf{r},t) &= \sum_i^{N_\mathrm{basis}} c_{in}(t) \phi_i\left(\mathbf{r}-\mathbf{R}_{I(i)}\right).
\end{align}
The expansion coefficients $c_{in}(t)\in\mathbb{C}$ here contain the time-dependence of the electronic system and are from now on expressed as a matrix $\mathbf{C}\in\mathbb{C}^{N_\mathrm{basis}\times N_\mathrm{occ}}$, indicating that only initially occupied orbitals are evolved in time. The time-dependent Kohn-Sham matrix equation 
\begin{align}
\frac{d}{dt}\mathbf{C}(t) &= -i\mathbf{S}^{-1}\mathbf{H}(t)\mathbf{C}(t)
\end{align} 
with the overlap matrix $\mathbf{S}\leftrightarrow \langle\phi_i|\phi_j\rangle$ and the Hamiltonian matrix $\mathbf{H}\leftrightarrow \langle\phi_i|\mathcal{H}^\mathrm{KS}|\phi_j\rangle$ is then to be solved to describe electron dynamics. The efficient and accurate solution of this equation is the key functionality in every RT-TDDFT code. We completely employ the highly optimized real-space integration framework to compute the density and Hamiltonian and overlap matrices as already existing in the code.\\~\\
Different approaches exist to solve the time-dependent KS equation, but for simplicity, we only discuss the Exponential Midpoint (EM) method here which belongs to the class of exponential integration schemes (or 'propagators') \cite{prop-castro04, prop-exp-varga}. The time-discretized propagation equation in the EM approach is given as
\begin{equation}
\label{eq:em}
\mathbf{C}(t+\Delta t) = \exp\Big(-i\Delta t\mathbf{S}^{-1}\mathbf{H}(t+\Delta t/2)\Big) \mathbf{C}(t)
\end{equation}
where $\Delta t$ is the integration time step. The matrix exponential can in our implementation be computed via eigenvectors \cite{mexp-moler}, defined as
\begin{equation}
\exp(\mathbf{M}) = \mathbf{V}_\mathbf{M}\,\mathrm{diag}\Big( e^{\lambda_1},...,e^{\lambda_n} \Big) \,\mathbf{V}_\mathbf{M}^{-1},
\end{equation}
where $\mathbf{V}_\mathbf{M}$ and $\lambda_i$ are the eigenvectors and -values of the matrix $\mathbf{M}\in \mathbb{C}^{n\times n}$, respectively. The advantage of this approach is that the highly optimized eigensolver functionality already built into the code can be used. We also implemented another method to compute the exponential, namely the so-called `Scaling and Squaring' approach based on the Padé approximation \cite{mexp-higham}.\\
The Hamiltonian matrix in eq. \ref{eq:em} is evaluated at one half time step in the future which makes this an implicit scheme. A common method to solve this type of equation is the Predictor-Corrector (PC) method \cite{numerical-recipes}. Here, a first guess based on the instantaneous Hamiltonian matrix is used to generate a predictor density which is then mixed with the initial density to perform a corrector step -- this can be repeated until convergence:
\begin{alignat*}{2}
& \mathrm{1.~Predictor:}~ \\
& ~~~~~~~~~~~ \mathbf{C}_p = \exp\Big(-i\Delta t\mathbf{S}^{-1}\mathbf{H}(t)\Big) \mathbf{C}(t) \\[3pt]
& \mathrm{2.~Update:} ~ \\
& ~~~~~~~~~~~ {\rho}_p = \rho(\mathbf{C}_p) \\[3pt]
& ~~~~~~~~~~~ \mathbf{H}_c = \frac{1}{2}\Big( \mathbf{H}[\rho(t),t] + \mathbf{H}[\rho_p, t+\Delta t] \Big) \\[3pt]
& \mathrm{3.~Corrector:}~ \\
& ~~~~~~~~~~~ \mathbf{C}(t+\Delta t) = \exp\Big(-i\Delta t\mathbf{S}^{-1}\mathbf{H}_c \Big) \mathbf{C}(t)
\end{alignat*}
 We also employ this approach for further discussions, but for completeness we note that we implemented several structurally different integration schemes in our code.

\section{Results}
The calculation of absorption spectra is a good way to test the stability and accuracy of a real-time TDDFT method because quite simple examples can be found for which both experimental (via absorption spectroscopy) and theoretical data (via linear-response TDDFT or more sophisticated approaches, e.g. based on the Bethe-Salpeter equation \cite{ bse-base}) exist. For this demonstration, we chose the Ethene (C$_2$H$_6$) molecule which is a frequently used testcase also incorporated in the popular test set of Thiel and coworkers \cite{thielbench1}.\\
The absorption spectrum can be calculated by RT-TDDFT via three individual calculations, each with the application of a weak delta-kick external field $\mathbf{E}(t) \sim \mathbf{E}_0\delta(t-t_0)$ along one of the cartesian axes. The Fourier transform of the response of the electronic dipole moment $\bm{\mu}(t)$ can then be used to calculate the polarization tensor $\alpha(\omega)$ which in turn is then used to calculate the absorption strength $S(\omega)$:
\begin{align}
\alpha_{ij}(\omega) &= \frac{ \mathcal{F}[\mu_i](\omega) }{ \mathcal{F}[E_j](\omega) } \\
S(\omega) &= \frac{2\omega}{3\pi}\mathrm{Im}\Big\{ \mathrm{Tr}[\alpha(\omega)] \Big\}
\end{align}
where $\mathcal{F}$ denotes the Fourier transform. The following calculations were performed for different numerical basis sets, namely for `light', `tight' and `tight+aug2', each increasing in size and with two additional diffuse Gaussian basis functions in the latter case which was shown to improve accuracy in benchmarking calculations for absorption spectra \cite{bseaimsbench}. The specific composition of the used basis sets is given in table \ref{tab:tab_basis}.
\begin{table}[H]
	\caption{Specifications for the `light', `tight' and `tight+aug2' basis sets used in our calculations. All basis functions are of hydrogenic type if not specified otherwise. Radial and angular quantum numbers indicate the shape, in parentheses are given the effective charges in the defining Coulomb potential. Gaussian functions are defined by gauss$_\mathrm{LN}$ where L specifies angular momentum and N the number of primitive Gaussians, the exponent (in $a_0^{-2}$) is given in the following parentheses. Each row adds another basis function to the last filled row above in that column.}
	\centering
	\begin{tabular}{lcc}
		\toprule
		& \multicolumn{2}{c}{Basis functions}                   \\
		\cmidrule(r){2-3}
		%Basis set     & $\Delta E_\mathrm{LR-TDDFT}^*$ (eV) & $\Delta E_\mathrm{RT-TDDFT}^*$ (eV) \\
		Basis set     & Hydrogen & Carbon \\
		\midrule
		`light'      & minimal   & minimal   \\
		             & 2s (2.10) & 2p (1.70) \\
		             & 2p (3.50) & 3d (6.00) \\
		             &           & 2s (4.90) \\
		`tight'      & 1s (0.85) & 4f (9.80) \\
				     & 2p (3.70) & 3p (5.20) \\
		             & 2s (1.20) & 3s (4.30)  \\
		             & 3d (7.00) & 5g (14.40) \\
		             &           & 3d (6.20)  \\
		`tight+aug2' & gauss$_{01}$ (0.02)  & gauss$_{01}$ (0.04)  \\
		             & gauss$_{11}$ (0.07)  & gauss$_{11}$ (0.03) \\
		\bottomrule
	\end{tabular}
	\label{tab:tab_basis}
\end{table}

For our RT-TDDFT calculations, we used the PBE exchange-correlation functional \cite{pbe}, a time-step of $0.1~\mathrm{a.u.}=0.0024~\mathrm{fs}$, a total simulation time of $2000~\mathrm{a.u.}=48.4~\mathrm{fs}$ and the EM propagator with the predictor-corrector solver. The external delta-kick field was described in the length gauge and defined by a Gaussian with a temporal width of $0.8~\mathrm{a.u.}=0.019~\mathrm{fs}$, centered at $t_0=50~\mathrm{a.u.}=1.2~\mathrm{fs}$ and with a maximum amplitude of $E_0 = 0.01~\mathrm{a.u.} = 0.51~$V/\AA. \\
For comparison, we performed LR-TDDFT calculations also with the FHI-aims code \cite{fhiaims}. The same pre-relaxed geometries (i.e. with PBE and the different basis sets as noted before) were used here and we employed the PW-LDA exchange-correlation kernel \cite{grosskohntdlda} for the calculations due to the lack of more sophistiacted kernels.

\begin{figure}[H]
	\centering
	\includegraphics[width=\linewidth]{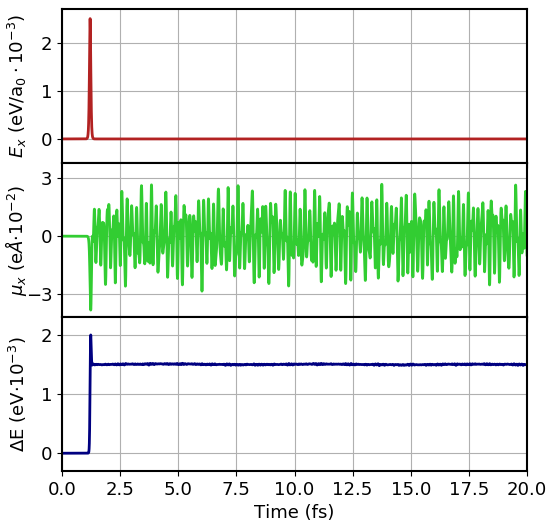}
	\caption{Time series results for the application of a delta pulse electric field along the x-axis on the Ethene molecule. Electric field amplitude (top row), electronic dipole response (mid row) and total electronic energy (bottom row).}
	\label{fig:fig1}
\end{figure}

\begin{figure}[H]
	\centering
	\includegraphics[width=\linewidth]{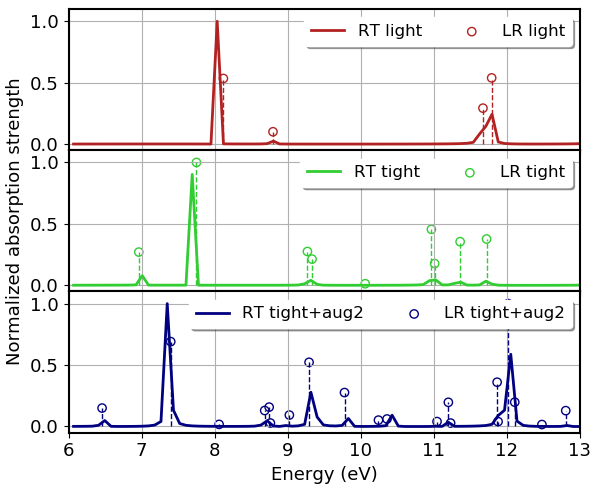}
	\caption{Absorption characteristics for different basis set sizes. Absorption spectra (for RT-TDDFT) and oscillator strengths (for LR-TDDFT) are normalized to respective maximum values. `Light' basis set (top row), `tight' basis set (mid row) and `tight+aug2' basis set (bottom row).}
	\label{fig:fig2}
\end{figure}

To illustrate the simulated dynamics, fig. \ref{fig:fig1} shows a time series for a calculation performed with a field oriented along the x-axis (parallel to the normal defining the molecule plane), which is shown in the first row (from the top). The second row shows the electronic dipole moment along the x-axis and one can clearly identify oscillations induced by the external field kick applied at around 1.2 fs. The bottom row shows the total time-dependent electronic energy where one can observe an increase after the delta pulse was applied -- this reflects the process of energy absorption of the electronic system from the external field which is not surprising due to the broad (i.e. constant) excitation spectrum. The constant energy after the pulse event demonstrates the excellent energy conservation on the below meV scale of our algorithm.\\
Figure \ref{fig:fig2} shows the calculated absorption characteristics obtained via RT-TDDFT (RT) and LR-TDDFT (LR) for the different basis sets `light', `tight' and `tight+aug2', where the latter denote the augmented `tight' basis set as defined in table \ref{tab:tab_basis}. The plots include the RT-TDDFT absorption spectra and the LR-TDDFT singlet oscillator strengths, each normalized with respect to the respective maxima on the whole spectrum. \\
It is obvious that the agreement between RT-TDDFT and LR-TDDFT is quite good in general. Every visible RT peak is associated with a closely located LR peak where also the relative magnitudes appear correlated. The spectrum generated with the `light' basis set shows only 3 (4) peaks for RT (LR) in the chosen range. The difference in the location of the large peak at 8 eV between RT and LR is around $\Delta E_\mathrm{max}^\mathrm{light}=0.08$ eV.\\
The `tight' basis set spectrum shows more structure as in the former case for both RT and LR. When comparing results between `light' and `tight' basis sets, a redshift is observed in the latter case which can most clearly be seen on the most prominent peak around 7.7 eV. The difference in the maximum peak location is here slightly reduced to $\Delta E_\mathrm{max}^\mathrm{tight}=0.06$ eV.\\

The same characteristic as seen in and between the `light' and `tight' results is visible for the `tight+aug2' data set, i.e. LR and RT spectra agree very well in the whole visible range, both also showing more structure, plus peaks being shifted. While several peaks in the range between 9 eV and 11 eV coincide with some in the `tight' spectrum, this is generally not the case. Another redshift is observed here at least for the first two peaks in the spectrum. The difference in the maximum peak located around 7.4 eV is again reduced by 0.02 eV to $ \Delta E_\mathrm{max}^\mathrm{tight+aug2}=0.04$ eV. This characteristic could point to a convergence of both approaches with increasing basis set size which should also be expected.

% discuss XC
% global blueshift
% conclude: RT = LR from tight on

%\end{multicols*}

To assess the overall accuracy of our RT-TDDFT implementation, we further provide a comparison with best estimated values, alongside with results obtained by LR-TDDFT. Table \ref{tab:table} shows the spectral location of the $1^1B_{1u}(\pi\rightarrow\pi^*)$ state transition peak, visible in fig. \ref{fig:fig2} as the large peaks between 7 eV and 8 eV, depending on the basis set size, relative to the `best estimate' literature value as defined by Thiel and coworkers \cite{thielbench1}, based on zero-point corrected experimental data \cite{davidson-98-ethene-vert}. While the reference energy is overestimated in our case for the `light' basis set and underestimated for the `tight+aug2` basis set, it still fits best for the `tight' basis set. Calculations done before with LR-TDDFT and BSE+G$_0$W$_0$ \cite{bseaimsbench}, both with the `tight+aug2' basis set, also show noticeable deviations of the lowest singlet excitation energy of Ethene for the same reference value, as observed here, too.

\begin{table}[H]
	\caption{Comparison of TDDFT vertical singlet excitation energies of the Ethene $1^1B_{1u}(\pi\rightarrow \pi^*)$ state relative to the best value (as defined in \cite{thielbench1}) of $E_\mathrm{ref}^*=7.80$ eV from \cite{davidson-98-ethene-vert}  and relative difference between LR-TDDFT and RT-TDDFT values.}
	\centering
	\begin{tabular}{lcc}
		\toprule
		%\cmidrule(r){1-2}
		Basis set     & $E_\mathrm{ref}^*-E_\mathrm{LR}$ (eV) & $E_\mathrm{ref}^*-E_\mathrm{RT}$ (eV) \\
		\midrule
		`light'      & -0.31  & -0.23   \\
		`tight'      & 0.02  & 0.11   \\
		`tight+aug2' & 0.41  & 0.45  \\
		\bottomrule
	\end{tabular}
	\label{tab:table}
\end{table}

At this point, we suspect that the close agreement of the `tight' basis set results with the reference value is fortuitous. A significant improvement could here probably be achieved by the use of hybrid functionals, e.g. the B3LYP functional \cite{becke-1993-b3lyp}, which are not yet available in our implementation.\\~\\
Regarding the computational cost of our simulations, we present the relative total simulation time scalings for both methods in table \ref{tab:tab-time}. 

\begin{table}[H]
	\caption{Total simulation time changes $\alpha=t_\mathrm{tot}/t_0$ relative to `light' basis reference simulations both for LR-TDDFT and RT-TDDFT. The reference base values are $t_{0,\mathrm{LR}}=1.5$ s and $t_{0,\mathrm{RT}}=28800.0$ s for LR- and RT-TDDFT, respectively. Note that the total simulation time for the Real-Time TDDFT case is the sum of the three individual calculations' times for the three cartesian field components. All calculations were carried out on a 4-core Intel Core i7 desktop machine, i.e. 4 MPI-parallel tasks via ScaLAPACK/ELSI \cite{elsi}.}
	\centering
	\begin{tabular}{lrr}
		\toprule
		%\cmidrule(r){1-2}
		Basis set     & $\alpha_\mathrm{LR}$ & $\alpha_\mathrm{RT}$ \\
		\midrule
		`light'      & 1.0  & 1.0   \\
		`tight'      & 12.8  & 4.1  \\
		`tight+aug2' & 190.93  & 37.3  \\
		\bottomrule
	\end{tabular}
	\label{tab:tab-time}
\end{table}

We do explicitly not attempt to compare the overall performance of both methods for the purpose of absorption spectra calculations, but rather shed light on the scaling of our method. Looking at the total computation time, it is obvious that LR-TDDFT is computationally much cheaper than RT-TDDFT for this task (nevertheless, the performance of RT-TDDFT can be optimized for this task by fitting analytically derived dipole response functions to a delta-like perturbation to the numerical dipole moment, enabling smaller propagation times due to increased accuracy, see \cite{kuemmel-rttddft-2018}).\\
Nevertheless, the overall scaling is quite different since the critial operations -- evaluating real-space double integrals plus solving a very large eigenvalue problem in case of LR-TDDFT and real-space evaluation of the electron density plus integration of the Hamiltonian matrix in case of RT-TDDFT -- have different scaling characteristics with basis set size $N_\mathrm{basis}$. \\
For most RT-TDDFT simulations, often $\mathcal{O}(10)$ atoms in a unit cell or molecular system, the grid-based operations, i.e. density update and Hamiltonian matrix integration, dominate the computational demand clearly, i.e. above $97\%$ in this case. The associated grid computation time (e.g. already observed in the initial SCF procedure) times the number of real-time steps thus determines the total simulation time. Using an implicit solver requires at least (and usually only) 2 of these grid operations which are already highly optimized. This makes our implementation very predictable and indicates very little unnecessary numerical overhead.

\section{Conclusion}
We have shown here the correctness of our Real-Time TDDFT implementation at least for the calculation of molecular absorption spectra, indicating its general validity for other applications, too. Good agreement of the optical absorption peaks with the already well-established LR-TDDFT functionality \cite{bseaimsbench} was achieved for all basis function sets, forming the basis of our argumentation. \\
An important insight for future applications is the dependence of the absorption peaks on the size of the basis set.\\
Further analysis of the influence of the chosen functional on the accuracy of the results could yield more insight and will be done in the future. \\
Regarding the computational performance, we have shown that our implementation has a clear scaling behaviour and that little numerical overhead is observed in the small molecule case. We expect this to be the general case due to the use of modern numerical methods. An analysis of the scaling with regard to computational resources will be done later. This also applies to an evaluation of the parameter space of the time-propagation framework, most importantly the time stepping and the choice of the propagation scheme.\\
Finally, we note that our code incorporates other interesting features, i.e. the possibility to perform Ehrenfest dynamics, various numerical options, e.g. propagation schemes, or also treatment of periodic systems -- all of this for a wide range of features already incorporated into the FHI-aims code, e.g. scalar-relativistic treatment.

\section*{Acknowledgements}

This work was funded by the Deutsche Forschungsgemeinschaft (DFT, German Research Foundation) via the CRC1242 `Non-Equilibrium Dynamics of Condensed Matter in the Time Domain'  (Project ID 278162697). The authors thank Volker Blum and Ville Havu for their support.

\bibliographystyle{unsrt}
\bibliography{ref}  

\end{multicols*}

\end{document}